\documentclass[preprint,showpacs,preprintnumbers,amsmath,amssymb]{revtex4}

\usepackage{graphicx}
\usepackage{dcolumn}
\usepackage{bm}

\begin{document}

\def\Journal#1#2#3#4{{#1} {\bf #2,} { #3} (#4)} 
 
\def\BiJ{ Biophys. J.}                 
\def\Bios{ Biosensors and Bioelectronics} 
\def\LNC{ Lett. Nuovo Cimento} 
\def\JCP{ J. Chem. Phys.} 
\def\JAP{ J. Appl. Phys.}
\def\JACS{ J. Am. Chem. Soc.} 
\def\JMB{ J. Mol. Biol.} 
\def\CMP{ Comm. Math. Phys.} 
\def\LMP{ Lett. Math. Phys.} 
\def\NLE{{ Nature Lett.}} 
\def\NPB{{ Nucl. Phys.} B} 
\def\PLA{{ Phys. Lett.}  A} 
\def\PLB{{ Phys. Lett.}  B} 
\def\PRL{ Phys. Rev. Lett.} 
\def\PRA{{ Phys. Rev.} A} 
\def\PRE{{ Phys. Rev.} E} 
\def\PRB{{ Phys. Rev.} B} 
\def\PNAS{{ Proc. Nat. Acad. Sci.}}
\def\EPL{{Europhys. Lett.} } 
\def\PD{{ Physica} D} 
\def\ZPC{{ Z. Phys.} C} 
\def\RMP{ Rev. Mod. Phys.} 
\def\EPJD{{ Eur. Phys. J.} D} 
\def\SAB{ Sens. Act. B} 
\title{Generalized Gumbel distribution of current fluctuations in purple membrane monolayers}

\author{E. Alfinito}
\thanks{eleonora.alfinito@unisalento.it}
\affiliation{Dipartimento di Ingegneria dell'Innovazione, Universit\`a del
Salento, via Monteroni, I-73100 Lecce-Italy (EU)}
\affiliation{CNISM - Consorzio Nazionale Interuniversitario per le Scienze Fisiche della Materia, via della Vasca Navale, 84, 00146  Roma-Italy (EU)
}%

\author{L. Reggiani}%
\thanks{lino.reggiani@unisalento.it}
\affiliation{Dipartimento di Matematica e Fisica "Ennio De Giorgi", Universit\`a del
Salento, via Monteroni, I-73100 Lecce-Italy (EU)}
\affiliation{CNISM - Consorzio Nazionale Interuniversitario per le Scienze Fisiche della Materia, via della Vasca Navale, 84, 00146  Roma-Italy (EU)
}%
%

\date{\today}
%
\begin{abstract}
We investigate the nature of a class of probability density functions, say $G(a)$, with $a$ the shape parameter, which generalizes the Gumbel distribution. These functions  appear in a model of charge transport, when applied to a metal-insulator-metal structure, where the insulator is constituted by a monolayer of bacteriorhodopsin. Current shows a sharp increase above about 3 V, interpreted as the cross-over between direct and injection sequential-tunneling regimes.
In particular, we show that, changing the bias value, the  probability density function changes its look from  bimodal to unimodal. Actually, the bimodal distributions can be resolved in at least a couple of  $G(a)$ functions with different  values of the shape parameter.
\end{abstract}
\maketitle

There is a widespread evidence that fluctuations of macroscopic observables exhibiting "extreme" events \cite{Cohen,Noullez,Bramwell02} can be described by means of the generalized Gumbel distribution $G(a)$, with $a$ the shape parameter:
\begin{equation}
G(a)=\frac{\theta(a)a^{a}}{\Gamma(a)}\exp\{-a[\theta(a)(z+\nu(a))+e^{-\theta(a)(z+\nu(a))}]\}
\label{eq:ga}
\end{equation}
%
%
where the function
 $\nu(a)$ is defined in terms of the Gamma function $\Gamma(a)$ and its derivatives
 $$\nu(a) =\frac{1}{\theta(a)}\left(\ln(a)-\psi(a)\right)$$
  $\Gamma(a)$, $\psi(a)$ and $\theta^{2}(a)$ indicating, respectively, the Gamma, digamma and trigamma function.
In the definition given in (\ref{eq:ga}), $G(a)$ is a normalized distribution function with zero mean and unitary variance.

This distribution is quite intriguing, since it crosses the perimeter of the extreme value statistics, describing physical phenomena of very different nature  \cite{Antal,Noullez,Bertin,Ciliberto}. 
Among its main features, here we recall that 
the well-known Bramwell-Holdsworth-Pintor (BHP)  distribution \cite{Bramwell}, found relevant for the description of critical
behaviors in magnetic, fluidodynamics \cite{Bramwell} and percolative systems \cite{PhysicaA} is practically coincident with  the distribution  $G(\pi/2)$ \cite{Noullez,Bertin,Ciliberto}.
Furthermore, for $a=1$ this distribution, also called \textit{scaled} Gumbel distribution, is used to describe the fluctuations of the conditional galaxy density \cite{Sylos}.
Finally,  for $a\rightarrow \infty$ the $G(a)$ recovers the Gaussian distribution.
Therefore, this distribution has a unifying character since it describes, on respect the value of $a$, the statistics of events from critical to non-critical conditions. 
\par
The shape parameter $a$ can take all the positive real numbers \cite{Bertin}.
The commonly accepted interpretation of integer valued $G(a)$ is the straightforward generalization of the Gumbel distribution, i.e. it describes the distribution of the $a$-th largest value of a set of independent and identically distributed (\textit{iid}) variables \cite{Noullez, Ciliberto}. 
Not so easy the interpretation of non-integer valued $G(a)$: in this case distributions are related to long, but finite, range correlations which do not allow for a single variable to be dominant \cite{Bertin}. 
In other words, when $a$ is not integer, $G(a)$ describes a finite-size system with correlations at all the scale lengths. 
As a rough distinction, we can say that integer $a$ values correspond to the establishment of ordering in the relevant variables, and non integer $a$ values correspond to a loss of ordering, a more democratic situation in which variance is large and the weight of events far from the mean is large.
\par
In this letter we investigate the fluctuations around the steady state of the current flowing through a monolayer of bacteriorhodopsin (bR)\cite{Oest}, an integral membrane protein sensitive to the light, as function of the applied voltage. 
By making use of an atomic force microscope (AFM) technique, in a large range of applied bias (up to about 8 V), it was experimentally observed \cite{Gomila} that the current exhibits a sharp transition between a near linear (Ohmic) regime and a  superlinear one (roughly as $V^8$ increase of current at a threshold voltage of about 3 V).
This behavior resembles a phase transition where the two different charge transport behaviors  can be associated with a  direct tunneling (DT) and an injection (or Fowler Nordheim) tunneling (FN) regime, respectively. 
This result was quantitatively reproduced within a microscopic model based on an impedance network protein analogous (INPA) \cite{PRE}. 
\par
\begin{figure}
       \centerline{
       \includegraphics[scale=0.4]{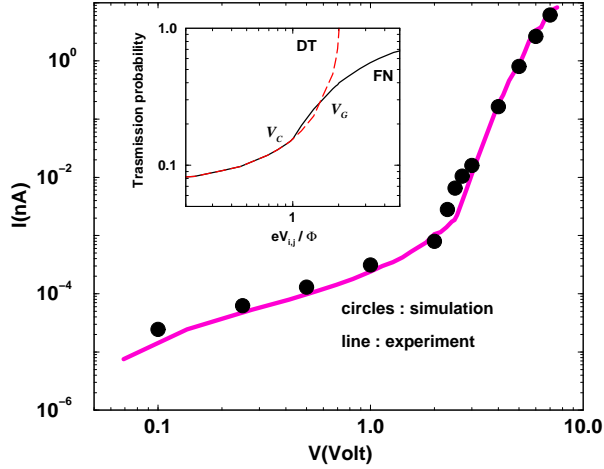}}
       \caption{Experimental and calculated data for the I-V characteristics of Ref.\cite{Gomila}.
       In the inset, the transmission probability as given by DT (dashed line) and the interpolation of DT and FN (continuous line) for the typical parameters: $m_e$ the free electron mass, $l_{i,j}= 5.5$  \AA, $\Phi$=219 meV. The critical, $V_C$, and Ginzburg $V_G$, voltages are indicated.}
       \label{fig:1}
\end{figure}

Figure \ref{fig:1} reports the current voltage (I-V) characteristic, as obtained by the experiments \cite{Gomila}(see continuous curve) and the theoretical model (see the full circles) when the AFM tip just touches the protein monolayer at about 4.6 nm from the bottom metal-contact.   
The transport model uses a stochastic approach to select the tunneling mechanism (DT or FN) based on the probability reported in the inset of Fig. 1.

By construction, the numerical approach allows for the simultaneous calculation of the current and its fluctuations around the steady state.
Accordingly, with respect to the steady value, the calculated current evolution is found to exhibit spikes that resemble "extreme" events, whose number increases  with the bias value, until it becomes difficult to establish what is extreme and what normal \cite{FNL}.
\par
The calculated variance of current fluctuations corresponding to the I-V characteristic of  Fig. (\ref{fig:1}), shows a 
a rather abrupt increase in concomitance with the cross-over region, at about 3 V.
The giant  increase, for about five orders in the magnitude of current variance, is associated with the opening of low resistance paths between contacts: they originate by the establishing of the FN regime which replaces the low voltage DT regime \cite{FNL}.
\par
%
%
\par
In the following we briefly survey the theoretical approach.
The INPA model aims to predict the static and dynamical electrical responses of a protein \textit{in vitro}, i.e. directly contacted  or sandwiched with functionalized contacts to an external bias. 
In particular, the single protein is described by means of an irregular impedance network, with ideal contacts \cite{PRE}.
The aminoacids constituting the tertiary structure of the protein are taken as the nodes of the network and the interaction between aminoacids as the links
of the network. 
For a given applied voltage the network is solved within Kirchhoff rules by associating an impedance to each link. 
The transfer of electrons between a couple of nodes, selected according to an interaction radius, is ruled by two different probabilities:
\begin{equation} 
\mathcal{P}^{D}_{i,j}= \exp \left[- \frac{2 l_{i,j}}{\hbar} \sqrt{2m_{e}(\Phi-\frac{1}{2}
eV_{i,j})} \right] \ ,
\hspace{0.7cm}
 eV_{i,j}  < \Phi  \,,
\label{eq:p1}
\end{equation}
for the DT regime and
\begin{equation}\label{eq:p2}
\mathcal{P}^{FN}_{ij}=\exp \left[-\left(\frac{2l_{i,j}\sqrt{2m_{e}}}{\hbar}\right)\frac{\Phi}{eV_{i,j}}\sqrt{\frac{\Phi}{2}} \right] \ , 
\hspace{0.7cm}
 eV_{i,j} \ge \Phi  \, ,
\end{equation}
for the FN regime.
Here $V_{i,j}$ is the local potential drop between the couple of $i,j$ amino-acids and $m_{e}$  is the electron effective mass, here taken the same of the bare value. 
A smooth variation of the aminoacid resistivity is introduced to take into account the superlinear current response:
\begin{equation}
\rho(V)=\left\{\begin{array}{lll}
\rho_{MAX}& \hspace{.5cm }& eV < \Phi  \\ \\
 \rho_{MAX} (\frac{\Phi}{eV})+\rho_{min}(1- \frac{\Phi}{eV}) &\hspace{.5cm} & eV \ge  \Phi 
 \end{array}
  \right.
\label{eq:3}
\end{equation}
where $\rho_{MAX}= 4 \times 10^{13} \ \Omega$ \AA \ is the resistivity value which should be used to fit the I-V characteristic at the lowest voltages,  $\rho_{min} = 4 \times 10^5 \ \Omega$ \AA \ plays the role of an extremely low series resistance, limiting the current at the highest voltages,  and $\Phi= 219$ meV  is the value of the  energy barrier separating two nodes, here taken to be the same for all the couples of nodes \cite{PRE}. 
\par
The electrical model well reproduces the measured I-V characteristic of bR.
On the other hand, since the two tunneling regimes are stochastically chosen according to their probability, the model provides also the instantaneous current fluctuations.
\par
\begin{figure}
        \centerline{
                \includegraphics[scale=0.45]{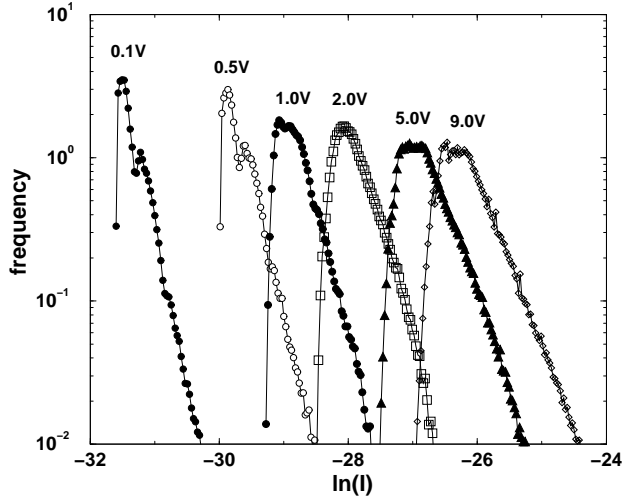}}
        \caption{Histograms of current fluctuations, V=0.1-9V.}
        \label{fig:3}
\end{figure}

To estimate the probability density functions (PDFs) of current fluctuations, we follow this procedure.
In a first step, we collect the histograms of ln(I) for different bias values in the range from 0.1  to 9 V. 
As reported in Fig. \ref{fig:3}, in all the bias range the histograms strongly deviate from a symmetric Gaussian-like shape.
Furthermore, at intermediate and high bias, i.e. near to the transition and beyond, they exhibit a nearly unimodal shape, while at bias lower than 1 V, the shape becomes bimodal.
\par
In a second step, we look for a fitting function of these histograms. 
For bias values larger than 1 V, the function:
\begin{equation}
H(a)= A_{0}\,\exp\left\{-a\, A_{1}(x+m)-a\, e^{-A_{1}(x+m)}\right\}
\label{eq:h}
\end{equation}
where $A_{0}, A_{1}, a, m$ are the curve parameters, gives a good approximation of calculated data. 
To reduce the histogram to a standard PDF, the above parameters should be adjusted. 
In doing so, we guess that the final PDF is the generalized Gumbel distribution (\ref{eq:ga}) with the same value of $a$; the standardization of histograms is performed by reshuffling the independent variable as follows:
$x=\sigma z+b$, and the new expression of the test function becomes: 
\begin{equation}
H'(z)=  A_{0}\,\exp\left\{-a\, (A_{1}\sigma)(z+\frac{b}{\sigma}+\frac{m}{\sigma}) -a\, e^{-A_{1}\sigma (z+\frac{b}{\sigma }+\frac{m}{\sigma})}\right\}
\end{equation}
which is proportional to $G(a)$, when assuming:
$$
\sigma=\frac{\theta_{a}}{A_1}, \,\, b=\left(\nu_{a}-\frac{m}{\sigma}\right)\sigma. 
$$  
In particular, $\lambda H'(z)=G(a)$ and the frequency normalization is
$$ \lambda(a)=\sigma\frac{a^a}{\Gamma(a)}\frac{A_{1}}{A_{0}} $$
The parameters $<x>, \ \sigma, \ \lambda$ are an $a$-generalization of the usual
mean value (location parameter) and the scale parameter\cite{wolfram}.
\par
\begin{figure}[htb]
 \centerline{
 \includegraphics[scale=0.4]{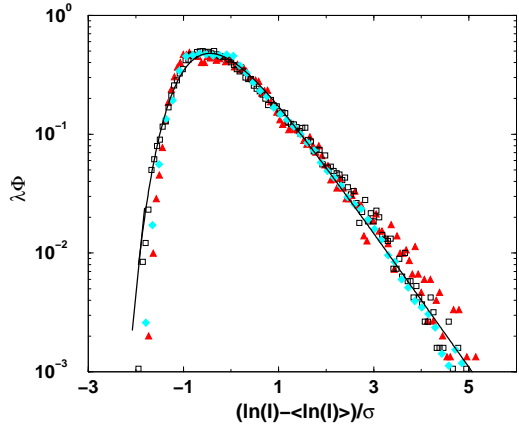}}
                \caption{Normalized distributions, V=1 V (triangles), 2 V (empty squares), 5 V(diamonds). The continuous curve is G(1).}
        \label{fig:4}
\end{figure}

In the region of bias values $1 \div 5 \ V$ this procedure gives a single PDF, the "scaled Gumbel" $G(1)$ \cite{Sylos} as shown in Fig. \ref{fig:4}. 
\par
\begin{figure}[htb]
\centerline{ \includegraphics[scale=0.4]{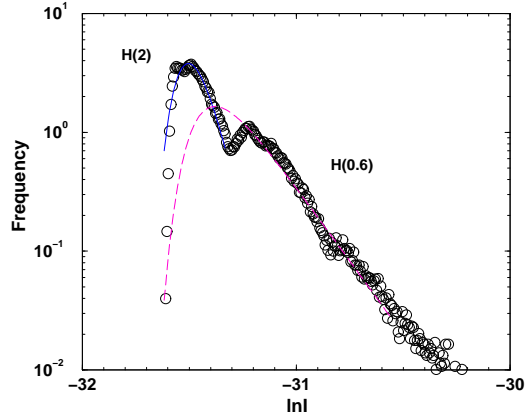}}
\caption{Histograms at V=0.1 V. Continuous line refers to the fitting function $H(2)$, dashed line refers to the fitting function $H(0.6)$, circles are the calculated data.}
\label{fig:5a}   
\end{figure}
\begin{figure}[htb]
 \centerline{
 \includegraphics[scale=0.4]{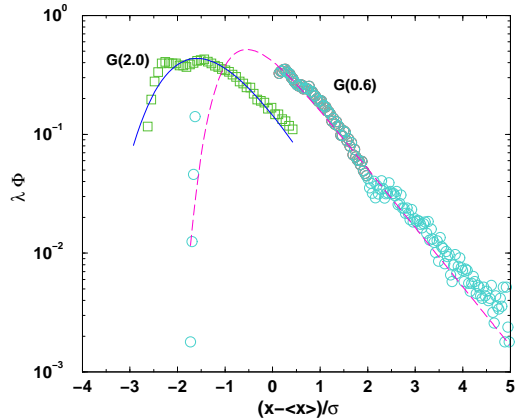}}
                \caption{PDFs with V=0.1 V. Continuous line refers to the fitting function $G(2)$, dashed line refers to the fitting function $G(0.6)$, symbols are the calculated data when rescaled (see text).}
        \label{fig:5b}        
\end{figure}

In a third step, we analyze the bimodal histograms. 
Following the suggestions of \cite{jask,Bramwell}, the bimodal shape is interpreted as the effect of the superposition of  different curves. 
In the present case we succeed in resolving two prevailing contributions: the $H(2)$-histogram and a $H(0.6)$-histogram.
By using the standardization technique shown above,  both $H(2)$ and $H(0.6)$ can be traced back to the $G(2)$ and $G(0.6)$ distributions, respectively. 
The comparison between the $H$s and $G$s curves is shown in Figs. \ref{fig:5a} and \ref{fig:5b}, for bias V=0.1 V. 
Notice that the curve on $G(2)$ could be also resolved in at least two different parts, thus signaling that the G(2) is a superposition of other PDFs. 
\par
\begin{figure}
        \centerline{
                \includegraphics[scale=0.4]{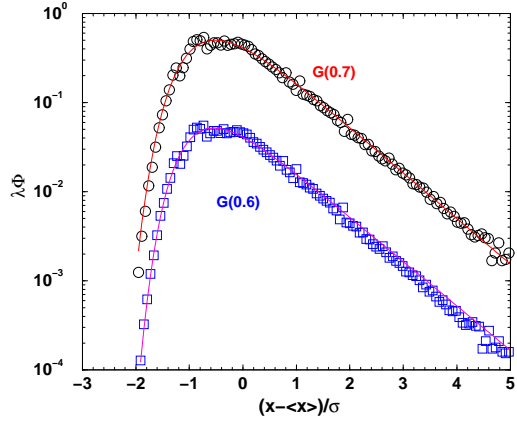}}
        \caption{PDFs of current fluctuations for V=9 V. Data have been compared with both the PDF $G(0.7)$ and $G(0.6)$. To better resolve the differences among the distributions, the $G(0.6)$ curve and the relative data are shifted of 0.1 on the vertical axis.}
        \label{fig:6}
\end{figure}

In the last step we observe that far from the transition, the $G(1)$  smooths down to a different unimodal distribution. 
Figure \ref{fig:6} reports  the calculated data and the fit with two different PDFs. 
The $G(0.7)$ gives the best fit.
\par
The most relevant result of the present investigation stems in solving the, apparently, substantial difference between the bimodal PDFs, evidenced before the transition, and the unimodal PDFs, emerging at the transition and beyond.
Bimodal distributions have been previously investigated as "mixture" of universal unimodal distributions \cite{jask}, but to our knowledge, a continuous transformation from bimodal to unimodal behaviour has not been observed.
A further contribution to this topic is the observation that the PDF:
\begin{equation}
G(a)=\frac{\theta(a)a^{a}}{\Gamma(a)}exp\{-aw-ae^{w}\}
\label{a1}
\end{equation}
is equivalent, in the sense of distribution, to the Gamma distribution of shape parameter $a$ and life-time $1/\lambda$:
\begin{equation}
f(t)=\frac{1}{\Gamma(a)} \lambda^{a}t^{a-1}e^{-\lambda t}.
\label{a2}
\end{equation}
This can be seen by using the change of variables:
$$ 
a e^{-\theta(a)w}=\lambda t
$$
and requiring the normalization \cite{Dufresne}.
\par
The Gamma distributions is well known in economics and granular materials \cite{Aste} and is usually interpreted as the distribution of a sum of $a$ (if $a$ is integer) $iid$ exponentially distributed variables characterized by the same life-time. 
In other words, the sum of different Gamma functions with the same life-time is still a Gamma function with a shape parameter equal to the sum of the single parameters \cite{Akkouchi}. 
Otherwise, for $iid$ with different life times, the variable sum (convolution) can be expressed by means of Gamma and Beta distribution functions \cite{Akkouchi} 
Therefore, we make the following conjecture: the $G(2)$ PDF we found in the low bias region, takes into account two different kinds of current, one due to the maximal resistivity (DT regime) and the other one due to the minimal resistivity (FN regime) \cite{FNL}.
The observed not-perfect convolution in a single $G(2)$ could be due to the (very) different life times of the events.
The $G(1)$ PDF signals the existence of a single dominant current regime, due to the superposition of all the possible resistivities values ($\rho(V), \ \rho_{MAX}, \ \rho_{min}$) (FN). 
This perfect superposition is announced by the $G(0.6)$ PDF which detaches by the $G(2)$ at low bias. 
The shape parameter lower than 1 signals the not-complete superposition. 
In addition, above the transition, the minimal resistivity becomes more and more dominant and this drains the $G(1)$ distribution producing a PDF with a shape parameter lower than 1, $G(0.7)$.

\par
The presence of $G(1)$ in a wide bias range suggests a further interpretation of the phenomenon: the strengthening of the system configuration. 
It happens that the phase transition between the two tunneling regimes is not abrupt but covers the bias value region 1$\div$ 5 V. 
In this region, at the microscopic level, the links can choose among the resistivity values $\rho_{MAX}$,  $\rho_{min}$ and $\rho(V)$ because both the DT and the FN regimes coexist. 
In different contexts, \cite{PRB2002}, the complete realization of the phase transition is associated with a temperature, called Ginzburg temperature, above which the ordering of the system is completed (in some cases with the formation of topological defects). 
By analogy, we can call the voltage value below which the transition is finalized (V=5 V), the Ginzburg voltage, $V_{G}$, and the voltage value at which the transition starts ($V_{C}$=1 \ V), the critical voltage $V_{C}$ (see Fig. \ref{fig:1}) \cite{FNL}  
\par
In conclusion, the main results presented in this letter are the following: 
\begin{enumerate}
        \item The probability density functions of current fluctuations show  both unimodal and bimodal shapes.
        \item The bimodal PDFs can be decomposed into at least two unimodal functions.
\item All the PDFs can be drawn back to the  parametric $G(a)$ distribution function(\ref{eq:ga}). 
\item The shape parameter $a$ is a function of the applied bias.
\end{enumerate}
\vspace{0.5cm}
\par
As final remark, we observe that what found on current fluctuations is a model prediction, since no measurements have been performed so far. 
The proposed model finely describe the I-V characteristics, and the interpretation of the PDF of fluctuations is in line with the theoretical mechanism of current transport leaving the experimental test as a future challenge. 
\begin{acknowledgments}
This research is supported by the European Commission under the Bioelectronic Olfactory Neuron Device (BOND) project within the grant agreement number 228685-2.
\end{acknowledgments}


\end{document}